\documentclass[12pt]{article}

\usepackage{sbc-template}
\usepackage{graphicx,url}
\usepackage[utf8]{inputenc}
\usepackage[english]{babel}
\usepackage{subcaption}
\usepackage{fontawesome}
\usepackage{amsmath} 
\usepackage{float}
\usepackage{acronym}
\usepackage{xcolor}
\usepackage{fancybox}
\usepackage{enumitem}
\usepackage{array} 
\usepackage{booktabs} 

\usepackage[absolute]{textpos}
\usepackage{tikz}

\title{Towards Cognitive Service Delivery on B5G through \\ AIaaS Architecture}

\author{Larissa Ferreira {Rodrigues Moreira}\inst{1,2}, Rodrigo Moreira\inst{1}, \\Flávio {de Oliveira Silva}\inst{2,3}, André Ricardo Backes\inst{4}}

\address{Federal University of Viçosa
  (UFV)\\
  Rio Paranaíba -- MG -- Brazil
\nextinstitute
  Faculty of Computing (FACOM) -- Federal University of Uberlândia (UFU)\\
  Uberlândia -- MG -- Brazil
\nextinstitute
  Department of Informatics -- School of Engineering\\
  University of Minho -- Braga, Portugal
\nextinstitute
  Department of Computing (DC) -- Federal University of São Carlos (UFSCar)\\
  São Carlos -- SP -- Brazil
  \email{\{larissa.f.rodrigues, rodrigo\}@ufv.br, flavio@di.uminho.pt,}  \email{\{larissarodrigues, flavio\}@ufu.br, arbackes@yahoo.com.br}
}

\begin{document} 
\acrodef{3GPP}{3rd Generation Partnership Project}
\acrodef{AI}{Artificial Intelligence}
\acrodef{AIaaS}{Artificial Intelligence as a Service}
\acrodef{ACM}{Association for Computing Machinery}
\acrodef{B5G}{Beyond Fifth Generation}
\acrodef{CUBIC}{Conjunctive Using BIC (Binary Increase Congestion Control)}
\acrodef{cwnd}{Congestion Window}
\acrodef{DoS}{Denial of Service}
\acrodef{DDoS}{Distributed Denial of Service}
\acrodef{DNN}{Deep Neural Network}
\acrodef{DRL}{Deep Reinforcement Learning}
\acrodef{DT}{Decision Tree}
\acrodef{DNN}{Deep Neural Network}
\acrodef{DMP} {Deep Multilayer Perceptron}
\acrodef{DQN}{Deep Q-Learning}
\acrodef{ETSI}{European Telecommunications Standards Institute}
\acrodef{eMBB}{enhanced Mobile Broadband}  
\acrodef{FL}{Federated Learning}
\acrodef{FIBRE}{Future Internet Brazilian Environment for Experimentation}
\acrodef{FTP}{File Transfer Protocol}
\acrodef{Flat}{Flat Neural Network}
\acrodef{GNN}{Graph Neural Networks}
\acrodef{HTM}{Hierarchical Temporal Memory}

\acrodef{IAM}{Identity And Access Management}
\acrodef{IEEE}{Institute of Electrical and Electronics Engineers}
\acrodef{IID}{Informally, Identically Distributed}
\acrodef{IoE}{Internet of Everything}
\acrodef{IoT}{Internet of Things}
\acrodef{IaaS}{Infrastructure as a Service}

\acrodef{KNN}{K-Nearest Neighbors}
\acrodef{KPI}{Key Performance Indicator}
\acrodef{KPIs}{Key Performance Indicators}
\acrodef{LSTM}{Long Short-Term Memory}
\acrodef{MPTCP}{Multipath Transmission Control Protocol}
\acrodef{M2M}{Machine to Machine}
\acrodef{MAE}{Mean Absolute Error}
\acrodef{ML}{Machine Learning}
\acrodef{MLaaS}{Machine Learning as a Service}
\acrodef{mMTC}{massive Machine Type Communications}
\acrodef{MOS}{Mean Opinion Score}
\acrodef{MAPE}{Mean Absolute Percentage Error}
\acrodef{MSE}{Mean Squared Error}
\acrodef{mMTC}{Massive Machine Type Communications}
\acrodef{MFA}{Multi-factor Authentication}
\acrodef{MQTT}{Message Queuing Telemetry Transport}

\acrodef{NN}{Deep Neural Network}
\acrodef{NS3}{Network Simulator 3}
\acrodef{NWDAF}{Network Data Analytics Function}
\acrodef{OSM}{Open Source MANO}
\acrodef{O-RAN}{Open Radio Access Network}
\acrodef{PAIaaS}{Pervasive Artificial Intelligence as a Service}
\acrodef{QL}{Q-learning}
\acrodef{QoE}{Quality of experience}
\acrodef{QoS}{Quality of Service}
\acrodef{OSS/BSS}{Operations Support System and Business Support System}
\acrodef{RAM}{Random-Access Memory}
\acrodef{RF}{Random Forest}
\acrodef{RL}{Reinforcement Learning}
\acrodef{RMSE}{Root Mean Square Error}
\acrodef{RNN}{Recurrent Neural Network}
\acrodef{Reno}{Regular NewReno}
\acrodef{RTT}{Round Trip Time}
\acrodef{SDN}{Software-Defined Networking}
\acrodef{SFI2}{Slicing Future Internet Infrastructures}
\acrodef{SLA}{Service-Level Agreement}
\acrodef{SON}{Self-Organizing Network}
\acrodef{SaaS}{Service as a Service}
\acrodef{SVM}{Support Vector Machine}

\acrodef{TCP}{Transmission Control Protocol}
\acrodef{URLLC}{Ultra Reliable Low Latency Communications}
\acrodef{VoD}{Video on Demand}
\acrodef{VNF}{Virtualized Network Function}
\acrodef{VR}{Virtual Reality}
\acrodef{V2X}{Vehicle-to-Everything}

\begin{textblock*}{15cm}(3cm,27.8cm) 
\noindent
    \footnotesize\textcolor{red}{This paper has been accepted by the WORKSHOP DE REDES 6G (W6G) 2024. 
    The definite version of this work was published by SBC-OpenLib as part of the W6G conference proceedings. 
    \\DOI: \url{https://doi.org/10.5753/w6g.2024.3304}}
\end{textblock*}
\maketitle

\begin{abstract}
Artificial Intelligence (AI) is pivotal in advancing mobile network systems by facilitating smart capabilities and automation. The transition from 4G to 5G has substantial implications for AI in consolidating a network predominantly geared towards business verticals. In this context, 3GPP has specified and introduced the Network Data Analytics Function (NWDAF) entity at the network's core to provide insights based on AI algorithms to benefit network orchestration. This paper proposes a framework for evolving NWDAF that presents the interfaces necessary to further empower the core network with AI capabilities B5G and 6G. In addition, we identify a set of research directions for realizing a distributed \textit{e}-NWDAF.

\end{abstract}

\section{Introduction}\label{sec:introduction}

To support different industry verticals, such as the Internet of Everything (IoE), Smart Factory, or Self Driving, many challenges and stringent requirements have been imposed on the standardization and development processes of 5G mobile networks~\cite{Coronado2022}. In particular, numerous advances have been made in the control plane of this network, such as network slicing~\cite{Moreira2021}, beamforming ~\cite{Brilhante2023}, closed-loop automation, \ac{AI}, and security enforcement~\cite{Silva2023} to support improved connectivity and chaining of virtualized network functions.

Different \ac{AI} techniques were applied to the core of the 5G network, aiming for efficient resource management and the delivery of connectivity to application vertices. Some disruptive techniques are built apart from the \ac{3GPP} specifications, while others are aligned with \ac{3GPP}, which is the case of the \ac{NWDAF} entity idealized in Release 15~\cite{3gpp2023}. The role of \ac{NWDAF} is to collect information from various network functions (NFs) and to support AI-based insights for the benefit of network management and orchestration~\cite{Ishteyaq2024}. However, the quality of the \ac{NWDAF} operation comes across all fields related to the availability of high-quality data; that is, the evolution of models depends on the data generated in the operator's domain, making generalization difficult~\cite{Gkonis2024}.

This paper aimed to design an evolution proposal aligned with \ac{3GPP} for \ac{NWDAF}, an analytics entity of a mobile 5G network core. The underlying concept behind our Evolved \ac{NWDAF} (\textit{e}-\ac{NWDAF}) is that it enables different players to access or offer third-party cognitive services in a mobile network through a northern interface on \ac{NWDAF}. In our vision, we present the interfaces, relationships, and behaviors for the realization of the evolved \textit{e}–\ac{NWDAF}. 

In addition, we discuss the \ac{AI} for Networking (AI4Net) paradigms as a way to enhance the \ac{NWDAF} functionalities of \ac{3GPP} and propose Networking for \ac{AI} (Net4AI) to deliver cognitive services at the core of mobile networks through the \ac{AIaaS}~\cite{Rodrigues2023} or NetApps~\cite{Slamnik2022} architectures. Our main contribution lies in an architectural framework to evolve \ac{NWDAF} to support different paradigms of offering cognitive services, as well as presenting a short survey with research directions towards evolving \ac{NWDAF}.

The remainder of this paper is organized as follows. Section~\ref{sec:related_work} surveys the related work, and Section~\ref{sec:proposed_method} presents the rationale of our approach. In Section~\ref{sec:research_directions}, we summarize a set of search directions around \ac{NWDAF} and present conclusions in Section~\ref{sec:concluding_remark}.

\section{Related Work}\label{sec:related_work}

This section provides a comparative review of state-of-the-art solutions that relate to both aspects of our proposal. The first refers to generic architectures for delivering cognitive services, while the other specifies approaches that target advancements in the core of the B5G network to enable \ac{AI} verticals.

\subsection{Generic AI Architectures}

\cite{Shah2022} proposed incorporating \ac{AIaaS}, \ac{SaaS} and \ac{IaaS} to curb the spread of unethical content on social media platforms. They proposed using AIaaS to identify and remove such content, which was classified into immoral, cyberbullying, and dislike categories. The authors employed traditional supervised learning algorithms but did not evaluate the effectiveness of deep learning techniques despite claiming that their approach is suitable for large datasets.

\cite{Fortuna2023}~investigated the potential benefits of using on-premise \ac{AIaaS} solutions for small and medium-sized businesses. The study evaluated the capabilities of \ac{AIaaS} and explored the feasibility of implementing it using open-source, user-friendly technologies. These technologies allow businesses to control their data, processes, and costs while minimizing reliance on third-party vendors or vendor lock-in risks.

\cite{Guntupalli2023} introduced a method to guarantee data compression, integrity, and confidentiality in \ac{AIaaS} deployments, which are online repositories that provide various \ac{ML} services and tools to users. However, selecting the best machine learning model or \ac{AI} service can be difficult. To address this challenge, \cite{Cerar2023} proposed a dynamic approach that utilizes \ac{DRL} to select the most appropriate \ac{ML} model and evaluate the method for a specific case of energy consumption forecasting.

\cite{Zhang2023} proposed an \ac{AIaaS} model deployment architecture for edge computing that enables the configuration of data quality and model complexity ratios for AI tasks, providing low-latency \ac{AI} capabilities. \cite{Hajipour2023} proposed a business plan incorporating AI products and services through AIaaS, featuring a strategic approach, step-by-step plan, and heuristic pricing model. However, the plan has not yet been implemented.

\cite{Baccour2023} proposed a novel platform architecture for deploying zero-touch \ac{PAIaaS} in 6G networks, supported by a blockchain-based smart system. They tested a use case for \ac{FL}, where the service consumer deployed distributed training on the 6G infrastructure with the assistance of the service provider's application, utilizing widespread devices.

Table~\ref{tab:related-work} summarizes the related works described above and highlights the differences demonstrating how our proposal represents a significant advancement in the field. The ``Native API for Edge'' column characterizes the solutions to offer native API in edge computing applications. The ``Multiple AI Facilities'' column characterizes the solutions to offer multiple \ac{AI} facilities such as different \ac{ML} algorithms, optimization methods, and feature extraction. Finally, the ``6G enabled'' column refers to 6G-enabled solutions.

\begin{table}[!htbp]
\caption{Short State-of-the-Art Survey.}
\scriptsize
\renewcommand{\arraystretch}{1.03}
\label{tab:related-work}
\resizebox{\textwidth}{!}{%
\begin{tabular}{lccc}
\hline
\multicolumn{1}{c}{\textbf{Approach}} & \textbf{Native API for Edge} & \textbf{Multiple AI Facilities} & \textbf{6G enabled} \\ \hline
\cite{Shah2022}                              & \faCircleO                    & \faCircle                        & \faCircleO           \\
\cite{Fortuna2023}                           & \faCircleO                    & \faCircle                        & \faCircleO           \\
\cite{Guntupalli2023}                        & \faCircleO                    & \faCircle                        & \faCircleO           \\
\cite{Cerar2023}                             & \faCircleO                    & \faCircle                        & \faCircleO           \\
\cite{Zhang2023}                             & \faCircleO                    & \faCircle                        & \faCircleO           \\
\cite{Hajipour2023}                          & \faCircleO                    & \faCircle                        & \faCircleO           \\
\cite{Baccour2023}                           & \faCircleO                    & \faCircleO                       & \faCircle           \\
\textbf{Our approach}                 & \faCircle            & \faCircle               & \faCircle   \\ \hline
\end{tabular}
}
\end{table}

\subsection{B5G-Enabled AI Architectures}

\cite{Li2021} proposed a cognitive service architecture for the 6G core network, inspired by the nervous system of an octopus, to enhance network performance and meet high service quality requirements. The cognitive service architecture aims to provide real-time perception, AI reasoning abilities, and a knowledge graph management system to adjust the core network dynamically. A case study is conducted to evaluate the performance of the cognitive service architecture, demonstrating significant time savings in session establishment compared to noncognitive service architecture. 

\cite{Manias2022} developed a functional prototype of NWDAF within a 5GC network, using unsupervised learning and clustering techniques to analyze intra-network interactions and optimize operations. The innovation lies in creating a working prototype for the 5GC network and NWDAF, enabling practical data-driven techniques and offering insights for future research and network management. 

\cite{Fiore2023} introduced a full network sensing paradigm, which systematically integrates sensing functionalities across all mobile network architecture domains. Their approach aims to unlock mobile networks’ untapped potential, allowing services to access rich metadata for knowledge discovery and informed policy-making. The authors highlight the need for a major shift in designing mobile networks for sensing, emphasizing the creation of new markets and innovative services that benefit society.

\cite{Hossain2023} introduced a distributed \ac{ML} approach using \ac{NWDAF} to optimize network operations. They presented two variations: one based on federated learning and the other on split learning, aiming to enhance data security, accuracy, and automation. The proposed architectures can greatly contribute to automating operations in 5G+ networks, yet challenges remain to be addressed.

\cite{Jeon2024} introduced a hierarchical network data analytics framework (H-NDAF), which distributes inference tasks to multiple leaf NWDAFs and centralizes training tasks at the root NWDAF, ensuring timely inference results while maintaining high accuracy. The H-NDAF was implemented using open-source software (free5GC) and evaluated through extensive simulations, particularly optimizing the policy for User Equipment (UE) data flows.

\section{Proposed Core Enhancement}\label{sec:proposed_method}

This paper proposes a background for \ac{NWDAF} enhancement Beyond 5G or 6G networks. The fundamental idea behind this enhancement is to extend the functions that \ac{3GPP} already designed in Release 18 for \ac{NWDAF} from two perspectives. The first perspective is to allow the network to benefit from \ac{AI} capabilities in a more sophisticated manner, with the operation being of the \ac{AI} for Networking (AI4Net) type. The second perspective is the return to enabling \ac{AI} functions for third parties through mobile networks, the operation being of the Networking for \ac{AI} (Net4AI) type.

The current standardization of \ac{NWDAF} provides analytics functions for mobile networks through a set of messages for data collection, reporting, subscription, and exposure of core entity metrics. \ac{NWDAF} allows for updating the pipeline with new analytics types, algorithms, or models and integrating them into operations. However, the \ac{NWDAF} standardization model is managed solely by network operators, hindering third-party innovation in delivering cognitive services. To address this, we propose combining and evolving \ac{NWDAF} with \ac{AIaaS} and NetApp concepts, currently in development, to specify how third-party services can be integrated and specified within the network framework.

\begin{figure}[!htbp]
  \centering
  \includegraphics[width=0.99\linewidth]{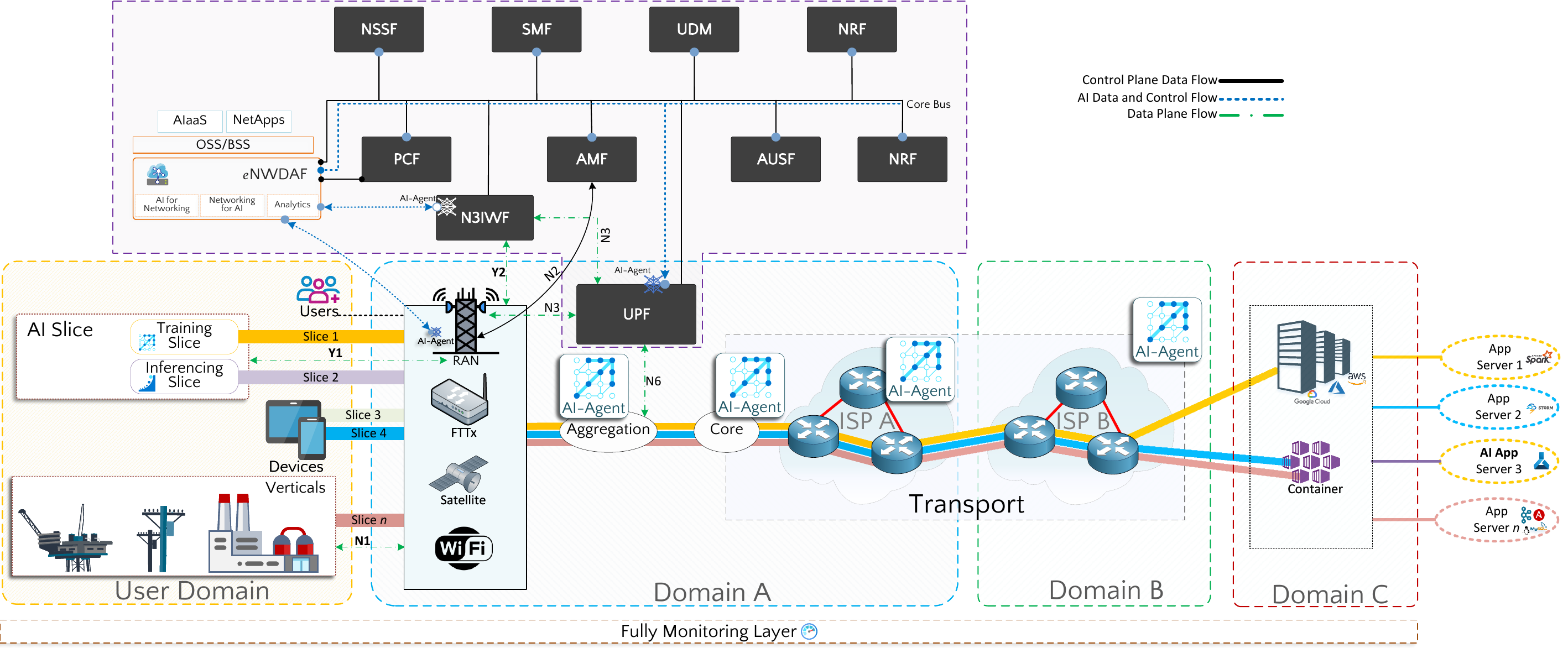}
  \caption{Evolved NWDAF to support AIaaS Cognitive Service Delivering.}
  \label{fig:method}
\end{figure}

\ac{AIaaS} refers to an architectural framework for lifecycle management and delivery of \ac{AI} services that aims to seamlessly deploy and embed intelligence capabilities to applications or edge devices~\cite {Rodrigues2023}. In line with \ac{ETSI}, the concept of NetApps was created, which is a 5G-enabled virtual application, and an evolution of the concept of \ac{VNF}, which provides functionalities to deliver complex services~\cite{Slamnik2022}. Thus, in Figure~\ref{fig:method}, we present the intervention proposed in this paper, which we call evolved \ac{NWDAF} (\textit{e}-\ac{NWDAF}).

In Figure~\ref{fig:method}, we position the main elements of the 5G core, and on the left, we highlight \textit{e}-\ac{NWDAF} with its additional functions: AI4Net and Net4AI. In our \textit{e}-\ac{NWDAF}, we idealize two evolution perspectives, namely, the \ac{OSS/BSS} interface, to enable the consumption and delivery of third-party cognitive services to users. The second refers to an architectural deployment model using \ac{AI}-Agents as microservices composed and distributed to each specific function of the network core.

The evolution perspective of \ac{NWDAF} aims to enable ``Training Slice'' or ``Inferencing Slice'' for users~\cite{Shen2022}. In addition to the potential to benefit network management at the core of the Service Provider, decoupling \ac{NWDAF} functionalities exclusively to the network operator and enabling the wake of new applications, such as Mobile Network Metrics as a Service. The second perspective of evolution enables a refined closed loop of \ac{AI} capabilities available to each core entity in the network.

With \textit{e}-\ac{NWDAF}, we envision performing network slices beyond connectivity, that is, network slices for training \ac{AI} or inference models. When idealizing 5G, a set of network requirements and business verticals would be enabled on mobile networks such as \ac{URLLC}, \ac{mMTC}, or \ac{eMBB}. For each of these verticals, the technical challenges of network slices as well as their \ac{KPIs} became known. With \textit{e}-\ac{NWDAF}, we envision a new vertical format for applications that can be enabled in B5G networks, such as NetAPPs or \ac{AIaaS} facilities.

\section{Research Directions}\label{sec:research_directions}

Towards distributed \ac{NWDAF} boosted by \ac{AI}, we conducted a short state-of-the-art analysis on the three main indexing databases, \ac{ACM}, ScienceDirect, and \ac{IEEE} Xplore, using the search string as below. The word combination provides a focused yet comprehensive approach to research, bridging network data analytics with advanced distributed learning techniques. This intersection aims brings practical challenges in data analysis and offers theoretical grounds to evolve \ac{AI} service composition.

\begin{center}
    \setlength{\fboxsep}{5pt}
    \shadowbox{\parbox{0.507\linewidth}{
        \footnotesize{\ttfamily{("NWDAF") AND ("Federated Learning")}}
    }}
\end{center}

We defined this search string because we are looking for approaches that assume that analytics functionalities at the core of the network can use distributed learning approaches such as federated learning. Analyzing these 35 papers, we sought to highlight and summarize the challenges and possible directions based on the notes of these authors. We only considered papers that contained consistent and practical approaches to \ac{NWDAF} and excluded short papers or surveys. In Table~\ref{tab:challenges_directions}, we summarize these challenges and highlight the research directions that should be addressed to realize or evolve the embedding intelligence in B5G systems.

\begin{table}[!htbp]
\scriptsize
\renewcommand{\arraystretch}{0.85}
\centering
\caption{Challenges and Possible Directions towards Realization and Evolving of \ac{NWDAF}.}
\label{tab:challenges_directions}
\resizebox{\textwidth}{!}{%
\begin{tabular}{ll} 
\toprule
\multicolumn{1}{c}{\textbf{Challenge}} & \multicolumn{1}{c}{\textbf{Possible Directions}}                                                                                                                                                                                                                                                                                                                                                                                                                                              \\ \bottomrule
Practical Deployment                   & \begin{tabular}{@{\labelitemi\hspace{\dimexpr\labelsep+0.5\tabcolsep}}l@{}}Efficient management and orchestration for space-aerial-terrestrial-ocean integrated.\\Intelligent and flexible management approaches for \ac{NWDAF}.\\Model and feature selection.\\Training acceleration.\\Containerization and interface design.\\Self-Coordination and self-organization of NWDAF leafs.\\Cross-domain training collaboration.\\Addressing statistical heterogeneity.\\AI/ML model maintenance.\end{tabular}  \\ \bottomrule
Scalability                            & \begin{tabular}{@{\labelitemi\hspace{\dimexpr\labelsep+0.5\tabcolsep}}l@{}}Interdomain scalability.\\Decentralization of NWDAF for real-time applications.\end{tabular}                                                                                                                                                                                                                                                                                                                       \\ \bottomrule
Privacy and Security                   & \begin{tabular}{@{\labelitemi\hspace{\dimexpr\labelsep+0.5\tabcolsep}}l@{}}Ensuring security and privacy protection on \ac{NWDAF} operation.\\Use FL to enable collaborative model training without disclosing private raw data.\\Trustworthiness of AI in \ac{NWDAF}.\\Security Overhead on Training and Inference life-cycle.\\Developing robust security measures that protect against intrusions\end{tabular}                                                                                                             \\ \bottomrule
AI Technologies                        & \begin{tabular}{@{\labelitemi\hspace{\dimexpr\labelsep+0.5\tabcolsep}}l@{}}AI/ML integration.\\Integrated FL services.\\Address the variance in FL update to reduce the effects of stragglers.~\\Cooperation and Sharing~ among multiple NWDAFs to train an ideal ML model.\end{tabular}                                                                                                                                                                                                      \\ \bottomrule
Communication                          & \begin{tabular}{@{\labelitemi\hspace{\dimexpr\labelsep+0.5\tabcolsep}}l@{}}Efficient protocols for \ac{NWDAF} message exchanging.\\Enhancing FL methods to minimize communication overhead.\end{tabular}                                                                                                                                                                                                                                                                                                                                \\ \bottomrule
AI Explainability                      & \begin{tabular}{@{\labelitemi\hspace{\dimexpr\labelsep+0.5\tabcolsep}}l@{}}Determining the reason of optimal placement of ML models on NWDAF leafs.\end{tabular}                                                                                                                                                                                                                                                                                                                                                                    \\ \bottomrule
Standardization~                       & \begin{tabular}{@{\labelitemi\hspace{\dimexpr\labelsep+0.5\tabcolsep}}l@{}}\ac{3GPP} and \ac{O-RAN} for AI/ML.\\Interoperability with other systems (e.g. \textit{non}-\ac{3GPP} systems).\end{tabular}                                                                                                                                                                                                                                                                                                                                       \\
\bottomrule
\end{tabular}
}
\end{table}

\section{Concluding Remarks}\label{sec:concluding_remark}

The present paper offers an evolutionary proposal for the \ac{NWDAF} system, which adopts a distributed architecture and employs \ac{AI}-Agents to interface with the various core functions of the mobile network. We undertook a comprehensive analysis of the core functions, identifying the aspects that are most suitable for enhancement following the specifications outlined by \ac{3GPP}. While \ac{NWDAF} in its original form represents a significant step forward in the field, it can be further refined to enable a broader range of benefits, particularly in cognitive service delivery.

We designed interfaces that enable third parties to receive cognitive services through Net4AI and expanded the fundamental principles of \ac{NWDAF} to encompass third-party cognitive services in the context of AI4Net. We identified research directions that focused on creating and verifying \textit{e}-\ac{NWDAF} architectures for B5G deployment. One of the limitations of this paper is the need for empirical validation of theoretical concepts. In future work, we plan to address this limitation by presenting validation and benchmark results.

\section*{Acknowledgments}

We thank CAPES and FAPESP MCTIC/CGI Research project 2018/23097-3 for the financial support. Andr\'e R. Backes gratefully acknowledges the financial support of CNPq (National Council for Scientific and Technological Development, Brazil) (Grant \#307100/2021-9). This study was financed in part by the Coordenação de Aperfeiçoamento de Pessoal de Nível Superior - Brasil (CAPES) - Finance Code 001. This work was also partially supported by RNP with resources from MCTIC, Grant No. 01245.010604/2020-14, under the Brazil 6G Project and Centro ALGORITMI, funded by Fundação para a Ciência e Tecnologia (FCT) within the RD Units Project Scope 2020-2023 (UIDB/00319/2020) for partially support this work.

\bibliographystyle{sbc}
\bibliography{Template_SBC/references}

\end{document}